\documentclass[twocolumn,aps,pra,amsmath,superscriptaddress,notitlepage,longbibliography]{revtex4-1}
\usepackage{amsmath,amssymb,amsfonts,bm}
\usepackage{graphicx}
\usepackage{dcolumn}
\usepackage{mathrsfs}
\usepackage{bbold}
\usepackage{dsfont}
\usepackage{dcolumn}
\usepackage[colorlinks=true,linkcolor=blue,citecolor=blue, urlcolor=blue,bookmarks=false]{hyperref}
\usepackage{changes}

\begin{document}


\title{Layer Hall effect induced by altermagnetism}

\author{Fang Qin}
\email{qinfang@just.edu.cn}
\affiliation{School of Science, Jiangsu University of Science and Technology, Zhenjiang, Jiangsu 212100, China}

\author{Rui Chen}
\email{chenr@hubu.edu.cn}
\affiliation{Department of Physics, Hubei University, Wuhan, Hubei 430062, China}

\begin{abstract}
In this work, we propose a scheme to realize the layer Hall effect in the ferromagnetic topological insulator Bi$_2$Se$_3$ via proximity to $d$-wave altermagnets. We show that an altermagnet and an in-plane magnetic field applied near one surface gap the corresponding Dirac cone, yielding an altermagnet-induced half-quantized Hall effect. When altermagnets with antiparallel N\'{e}el vectors are placed near the top and bottom surfaces, giving rise to the layer Hall effect with vanishing net Hall conductance, i.e., the altermagnet-induced layer Hall effect. In contrast, altermagnets with parallel N\'{e}el vectors lead to a quantized Chern insulating state, i.e., the altermagnet-induced anomalous Hall effect. We further analyze the dependence of the Hall conductance on the orientation of the in-plane magnetic field and demonstrate that the layer Hall effect becomes observable under a perpendicular electric field. Our results establish a route to engineer altermagnet-induced topological phases in ferromagnetic topological insulators.
\end{abstract}
\maketitle

\section{Introduction}

The layer Hall effect describes a distinctive electronic response in which charge carriers are spontaneously deflected toward opposite transverse sides in different atomic layers~\cite{gao2021layer,chen2024layer,anirban2023quantum,dai2022quantum,peng2023intrinsic,xu2024layer,lei2023kerr,tao2024layer,yi2024disorder,zhang2023layer,zhang2024layer,feng2023layer,liu2024engineering,han2025layer}.
This phenomenon has been experimentally observed in even-layered antiferromagnetic topological insulators such as MnBi$_2$Te$_4$~\cite{gao2021layer}, marking a major milestone in the exploration of layer-resolved Hall responses.
Theoretically, the concept has been extended to a variety of material platforms, including MnBi$_2$Te$_4$, In$_2$Se$_3$, and In$_2$Te$_3$ heterostructures~\cite{gao2021layer,chen2024layer,anirban2023quantum,dai2022quantum,peng2023intrinsic,xu2024layer,lei2023kerr}, transition-metal oxides~\cite{tao2024layer}, magnetic sandwich heterostructures~\cite{yi2024disorder}, valleytronic van der Waals bilayers~\cite{zhang2023layer}, inversion-symmetric monolayers~\cite{zhang2024layer}, and multiferroic materials~\cite{feng2023layer,liu2024engineering}.

In most cases, the layer Hall effect arises from an externally applied electric field~\cite{gao2021layer,chen2024layer,dai2022quantum,yi2024disorder,zhang2023layer,tao2024layer}, although similar responses can also be induced by internal electric fields generated through ferroelectric polarization~\cite{xu2024layer,liu2024engineering} or interlayer sliding~\cite{peng2023intrinsic,zhang2023layer,feng2023layer}.
Beyond electric-field-driven mechanisms, a distinct route has been proposed in which inequivalent exchange fields applied to the top and bottom surfaces of a topological-insulator thin film produce a spontaneous layer Hall effect, even without external bias~\cite{han2025layer}.
Importantly, the layer Hall effect has also been recognized as a key experimental signature of the axion insulator phase~\cite{yi2024disorder,wang2015quantized,morimoto2015topological,mogi2017magnetic,mogi2017tailoring,varnava2018surfaces,xiao2018realization,xu2019higher,zhang2019topological,liu2020robust,nenno2020axion,li2024dissipationless,qin2023light,li2024high}, providing a crucial connection between magnetic symmetry breaking and topological electromagnetic responses.

Altermagnetism represents a recently identified class of collinear magnetic phases distinguished by unique spin-group symmetries~\cite{wu2004dynamic,wu2007fermi,lee2009theory,yuan2026unconventional,hayami2019momentum,hayami2020bottom,smejkal2022giant,smejkal2022beyond,smejkal2022emerging,li2025exploration}.
Altermagnets exhibit anisotropic spin-split electronic bands and alternating collinear magnetic moments on adjacent lattice sites, setting them apart from conventional ferromagnets and antiferromagnets~\cite{wu2004dynamic,wu2007fermi,hayami2019momentum,hayami2020bottom,smejkal2022giant,smejkal2022beyond,smejkal2022emerging,li2025exploration}.
A rapidly expanding list of candidate altermagnetic materials includes RuO$_2$~\cite{li2025exploration,ahn2019antiferromagnetism,vsmejkal2020crystal,shao2021spin,rafael2021efficient,bose2022tilted,bai2022observation,karube2022observation,he2025evidence}, RuF$_4$~\cite{milivojevic2024interplay}, ReO$_2$~\cite{chakraborty2024strain}, MnF$_2$~\cite{bhowal2024ferroically,li2024creation}, FeSb$_2$~\cite{mazin2021prediction,attias2024intrinsic,phillips2025electronic}, CrSb~\cite{reimers2024direct,ding2024large,peng2025scaling,zhou2025manipulation},  MnTe~\cite{mazin2023altermagnetism,krempasky2024altermagnetic,lee2024broken,osumi2024observation,orlova2025magnetocaloric}, Mn$_5$Si$_3$~\cite{leiviska2024anisotropy,reichlova2024observation,rial2024altermagnetic}, (Ca,Ce)MnO$_3$~\cite{vistoli2019giant,fernandes2024topological},  KV$_2$Se$_2$O~\cite{jiang2025metallic,sarkar2025altermagnet}, and BiFeO$_3$~\cite{urru2025g,george2026topological,sajid2026anisotropic,gui2026electric}.

A variety of experimental schemes have been proposed to detect altermagnetism~\cite{lin2025coulomb,chen2025electrical,chen2025probing}.
Coulomb drag has been identified as a possible probe of altermagnetic order~\cite{lin2025coulomb}, while electrical switching of altermagnetic states has been theoretically demonstrated~\cite{chen2025electrical}.
Experimental setups for directly probing momentum-space spin polarization~\cite{chen2025probing} and distinguishing intrinsic from extrinsic spin-orbital altermagnetism via spin conductivity and orbital polarization~\cite{wang2025spin} have also been suggested.

Theoretical studies have revealed a wide range of unconventional phenomena in altermagnets, including the Josephson effect~\cite{ouassou2023dc,zhang2024finite,cheng2024orientation,beenakker2023phase,lu2024varphi,sun2025tunable,fukaya2025josephson}, Andreev reflection~\cite{sun2023andreev,papaj2023andreev}, nonlinear transport~\cite{fang2024quantum,liu2025enhancement}, magnetoresistance effect~\cite{sun2025tunneling}, parity anomaly~\cite{wan2025altermagnetism}, helical edge states~\cite{wan2025interplay},  quasicrystals~\cite{chen2025quasicrystalline,shao2025classification,li2025unconventional}, N{\'e}el spin currents~\cite{shao2023neel}, and thermoelectric Hall effect~\cite{qin2026anomalous}.
Additionally, numerous topological effects have been predicted, such as altermagnet-induced topological phases~\cite{rao2024tunable,ma2024altermagnetic,antonenko2025mirror,qu2025altermagnetic,parshukov2025topological,fernandes2024topological}, higher-order topological states~\cite{li2024creation}, floating edge bands~\cite{li2025floating}, light-induced odd-parity altermagnetism~\cite{zhuang2025odd,huang2025light,zhu2025floquet,liu2025light}, Floquet-engineered topological phases~\cite{zhu2025floquet_Chen}, and topological superconductivity~\cite{fu2025altermagnetism}.
These developments position altermagnetism as a promising platform for engineering and controlling topological and correlated quantum phases.

\begin{figure}[htpb]
\centering
\includegraphics[width=\columnwidth]{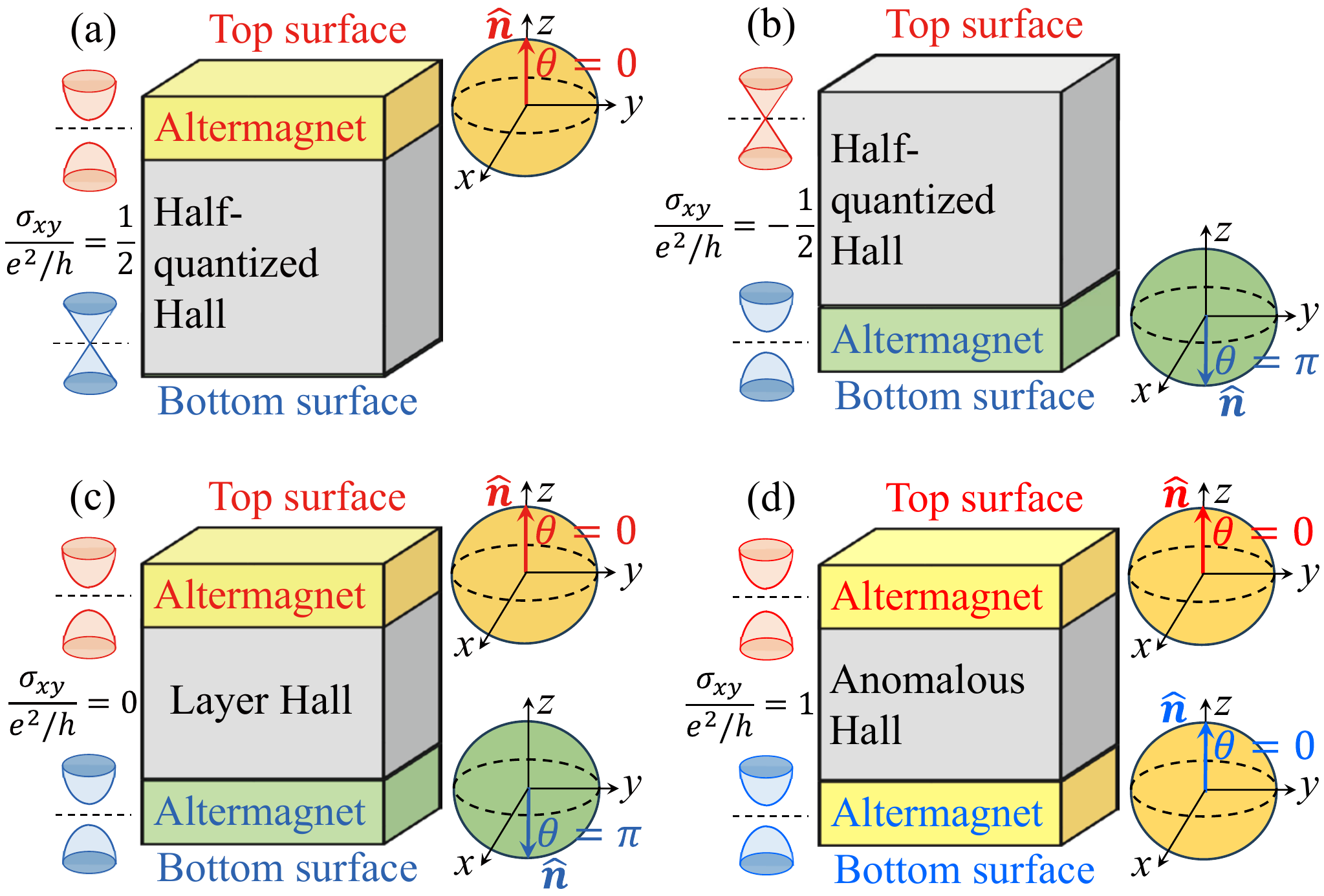}
\caption{Schematic of topological phases in a three-dimensional topological insulator Bi$_2$Se$_3$ in the presence of $d$-wave altermagnetic order on the top or bottom layers. The unit vector $\hat{\bf n}=(\sin\theta\cos\phi,\sin\theta\sin\phi,\cos\theta)$ denotes the orientation of the N\'{e}el vector~\cite{shao2021spin,li2024creation}, where $\theta$ and $\phi$ are the polar and azimuthal angles, respectively. For simplicity, $\theta\!=\!0$ or $\pi$.
(a) Altermagnet penetrating only the top layers (yellow) gaps the top surface Dirac cone via time-reversal-symmetry breaking, giving rise to an altermagnet-induced half-quantized Hall effect.
(b) Altermagnet penetrating only the bottom layers (green) gaps the bottom surface Dirac cone, leading to a half-quantized Hall effect.
(c) Antiparallel N\'{e}el vectors on the top and bottom surfaces gap both Dirac cones with opposite Hall contributions, resulting in an altermagnet-induced layer Hall effect with zero net Hall conductance.
(d) Parallel N\'{e}el vectors on the two surfaces gap both Dirac cones with identical Hall contributions, yielding an altermagnet-induced anomalous Hall effect and a fully quantized Chern insulating phase.} \label{Fig:Schematic_Hall_abcd}
\end{figure}

In this work, we propose a scheme to realize the layer Hall effect in the ferromagnetic topological insulator Bi$_2$Se$_3$ via proximity to $d$-wave altermagnets.
We consider a three-dimensional (3D) ferromagnetic topological insulator coupled to altermagnetic layers under an external in-plane layer magnetic field.
We find that inserting an altermagnet and applying an in-plane magnetic field to layers near either the top or bottom surface opens a gap in the corresponding surface Dirac cone, giving rise to an altermagnet-induced half-quantized Hall effect as shown in Figs.~\ref{Fig:Schematic_Hall_abcd}(a) and \ref{Fig:Schematic_Hall_abcd}(b). Specifically:
\begin{itemize}
\item When altermagnets with antiparallel N\'{e}el vectors~\cite{shao2021spin,li2024creation} are placed near the top and bottom surfaces, both Dirac cones become gapped with opposite Hall contributions, producing the altermagnet-induced layer Hall effect with vanishing net Hall conductance as shown in Fig.~\ref{Fig:Schematic_Hall_abcd}(c).
\item When altermagnets with parallel N\'{e}el vectors are applied to both surfaces, the two Dirac cones become gapped with the same Dirac mass, yielding a fully quantized Chern insulating state, i.e., the altermagnet-induced anomalous Hall effect as shown in Fig.~\ref{Fig:Schematic_Hall_abcd}(d).
\end{itemize}

Furthermore, we investigate the dependence of the Hall conductance on the orientation of the in-plane magnetic field and demonstrate that the layer Hall effect becomes experimentally accessible under a perpendicular electric field.
Our results establish a versatile strategy for realizing altermagnet-induced topological phases in ferromagnetic topological insulators, paving the way toward the design and implementation of altermagnet-based topological materials.

The remainder of the paper is organized as follows.
Section~\ref{2} introduces the model Hamiltonian for the 3D $\mathbb{Z}_{2}$ topological insulator Bi$_2$Se$_3$ coupled to altermagnetic layers under an in-plane layer magnetic field.
Section~\ref{3} presents numerical results for $d$-wave altermagnets.
In Section~\ref{4}, we derive the effective surface Hamiltonians and analytically obtain the Hall conductances.
In Section~\ref{5}, we show how a perpendicular electric field can be used to reveal the layer Hall effect.
Finally, Section~\ref{6} summarizes the main conclusions.

\section{Model}\label{2}

We investigate a 3D $\mathbb{Z}_{2}$ topological insulator, Bi$_{2}$Se$_{3}$, placed in proximity to altermagnetic layers and subjected to an external in-plane layer magnetic field. The corresponding low-energy effective Hamiltonian is written as
\begin{eqnarray}\label{eq:Hk_original}
\hat{\cal H}({\bf k}) \!=\! \hat{\cal H}_{\rm TI}^{}({\bf k}) \!+\! \hat{\cal H}_{\Delta}^{} \!+\! \hat{\cal H}_{J}^{}({\bf k}_{||}),
\end{eqnarray} where ${\bf k}\!=\!(k_{x},k_{y},k_{z})$ and ${\bf k}_{||}\!=\!(k_{x},k_{y})$. The three terms represent, respectively, the bulk Hamiltonian of Bi$_{2}$Se$_{3}$, the Zeeman-type spin splitting induced by magnetic doping, and the momentum-dependent altermagnetic exchange coupling arising from proximity to altermagnetic layers.

The first term, $\hat{\cal H}_{\rm TI}({\bf k})$, describes the bulk electronic structure of Bi$_{2}$Se$_{3}$ and takes the standard form~\cite{liu2010model}
\begin{eqnarray}
\hat{\cal H}_{\rm TI}({\bf k}) &\!=\!& {\cal M}({\bf k})\sigma_{0}^{}\otimes\tau_{z} \!+\! A_{1}k_{z}\sigma_{0}^{}\otimes\tau_{y} \nonumber\\ 
&&\!+ A_{2}(k_{y}\sigma_{x} \!-\! k_{x}\sigma_{y})\otimes\tau_{x}, \label{eq:Hk_TI}
\end{eqnarray}
where ${\cal M}({\bf k}) \!=\! M \!-\! B_{1}k_{z}^{2} \!-\! B_{2}(k_{x}^{2} \!+\! k_{y}^{2})$, and $\sigma_{x,y,z}$ ($\tau_{x,y,z}$) are Pauli matrices acting on spin (orbital) degrees of freedom. The material parameters are chosen as $M\!=\!0.28$ eV, $A_{1}\!=\!0.22$ eV$\cdot$nm, $A_{2}\!=\!0.41$ eV$\cdot$nm, $B_{1}\!=\!0.10$ eV$\cdot$nm$^{2}$, and $B_{2}\!=\!0.566$ eV$\cdot$nm$^{2}$, consistent with the well-established model of Bi$_{2}$Se$_{3}$~\cite{zhang2009topological,liu2010model,chang2013experimental,mogi2022experimental,zou2023half,qin2023light,qin2022phase,chen2025probing}.

The second term, $\hat{\cal H}_{\Delta}^{}\!=\!F(z)(\Delta_{x}\sigma_{y} \!-\! \Delta_{y}\sigma_{x})\otimes\tau_{0}^{}$, represents a Zeeman-type spin splitting arising from the exchange field induced by magnetic doping~\cite{liu2010model,zyuzin2020plane}. The in-plane spin splitting is primarily determined by the exchange interaction between electron spins and magnetic dopants, whose moments are aligned by the applied in-plane magnetic field. Consequently, the resulting exchange-driven splitting is substantially stronger than the direct Zeeman coupling arising from the field itself~\cite{chen2025probing,chu2011surface,liu2013in,fu2009hexagonal}. Here, $\Delta_{x}\!=\!\Delta\cos\varphi$ and $\Delta_{y}\!=\!\Delta\sin\varphi$, where $\Delta$ denotes the magnitude of the exchange field~\cite{yu2010quantized,chang2013experimental,kandala2015giant} and $\varphi$ specifies the in-plane orientation angle of the applied magnetic field. The function $F(z)$ captures the spatial profile of the magnetic proximity effect along the $z$ direction, ensuring that the induced magnetization is localized near the surface or interface~\cite{mogi2022experimental}.

The third term, $\hat{\cal H}_{J}^{}({\bf k}_{||})\!=\!G(z)J(k_{x},k_{y})\sigma_{z}\otimes\tau_{0}^{}$, 
describes the contribution from altermagnetic ordering induced by the adjacent altermagnetic layers. The function $J(k_{x},k_{y})$ encodes the momentum-dependent form factor associated with the altermagnetic order, while $G(z)$ characterizes the spatial dependence of this interfacial coupling.
Furthermore, we consider a $d$-wave altermagnetic term of the form $\hat{\cal H}_{J}^{}({\bf k}_{||})\!=\!G(z)J_{d}(k_{y}^{2} \!-\! k_{x}^{2})(\boldsymbol{\sigma}\cdot\hat{\bf n})\otimes\tau_{0}^{}\!=\!G(z)J_{d}(k_{y}^{2} \!-\! k_{x}^{2})\cos\theta\sigma_{z}\otimes\tau_{0}^{}$~\cite{smejkal2022giant,smejkal2022beyond,smejkal2022emerging}, where the unit vector $\hat{\bf n}\!=\!(\sin\theta\cos\phi,\sin\theta\sin\phi,\cos\theta)$ denotes the direction of the N\'{e}el vector~\cite{li2024creation}, with $\theta$ and $\phi$ being the polar and azimuthal angles in spherical coordinates, respectively. To realize the layer Hall effect, we set $\theta\!=\!0$ for the top surface states and $\theta\!=\!\pi$ for the bottom surface states, as illustrated in Fig.~\ref{Fig:Schematic_Hall_abcd}(c). In contrast, the anomalous Hall effect is obtained by choosing $\theta\!=\!0$ for both the top and bottom surface states, as shown in Fig.~\ref{Fig:Schematic_Hall_abcd}(d).

Together, these three components capture the essential physics of a ferromagnetic topological insulator in proximity to altermagnetic layers, providing the basis for exploring the emergence of the altermagnet-induced layer Hall effect.

\section{Numerical results}\label{3}

\begin{figure*}[htpb]
\centering
\includegraphics[width=\textwidth]{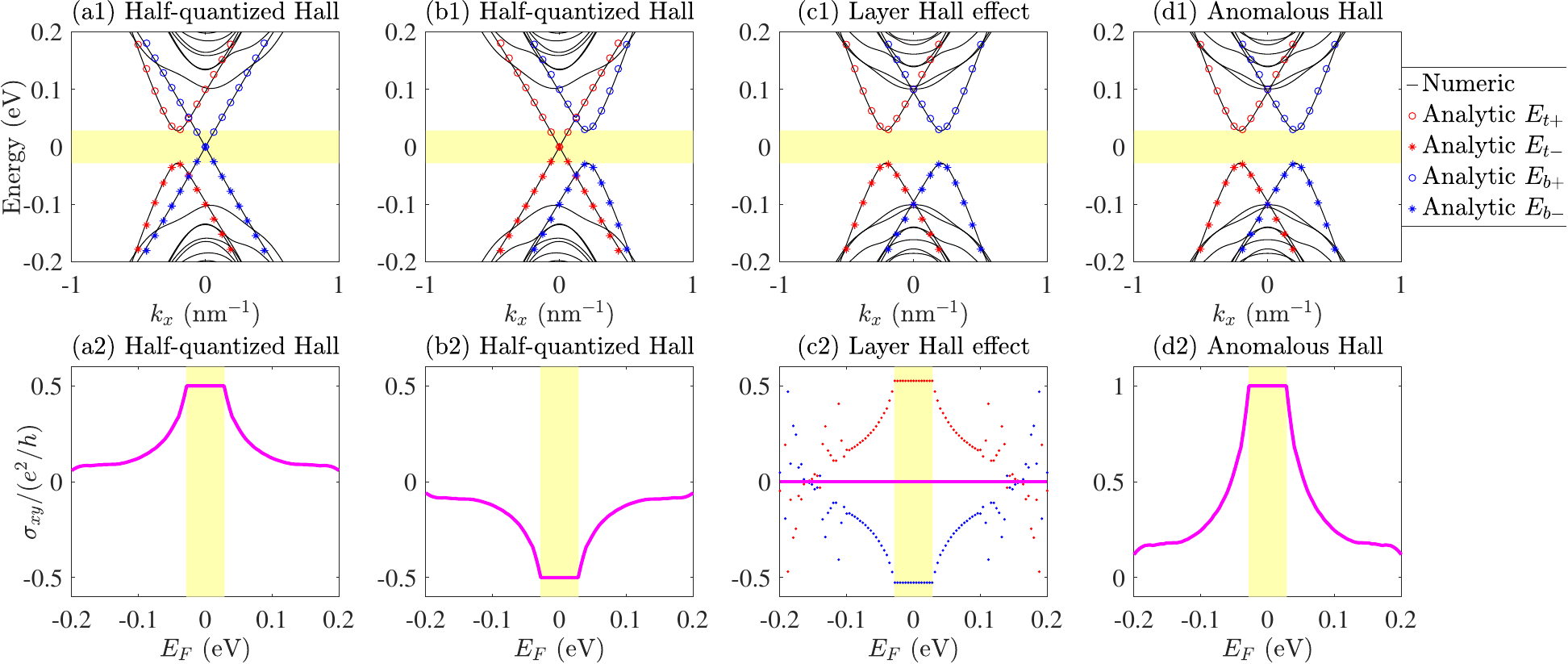}
\caption{Band structures and Hall conductances for the tight-binding Hamiltonian \eqref{eq:Hz_TB_r} of Bi$_2$Se$_3$ in proximity to $d$-wave altermagnetic layers under an external in-plane magnetic field.
[(a1)-(d1)] Energy spectra. Black curves show numerical results obtained under OBCs along $z$ direction and PBCs along $x$ and $y$ directions. Red (blue) circles and asterisks denote the analytical surface-state spectra for the top (bottom) surface, derived from Eqs.~\eqref{eq:Ek_top} and \eqref{eq:Ek_bot}. Subscripts label the surface bands: $t+$ (top conduction), $t-$ (top valence), $b+$ (bottom conduction), and $b-$ (bottom valence).
(a1) Top-surface altermagnetic layer with thickness $d_{t}\!=\!2$ nm, which gaps the top Dirac cone. Here, $\theta\!=\!0$, $G(z)\cos0\!=\!F(z)\!=\!1$ for $0\!\leqslant\!z\!\leqslant\!2$ nm and $G(z)\!=\!F(z)\!=\!0$ elsewhere.
(b1) Bottom-surface inverse altermagnetic layer with thickness $d_{b}\!=\!2$ nm, which gaps the bottom Dirac cone. Here, $\theta\!=\!\pi$, $G(z)\cos\pi\!=\!-F(z)\!=\!-1$ for $8\!\leqslant\!z\!\leqslant\!10$ nm and $G(z)\!=\!F(z)\!=\!0$ elsewhere.
(c1) Combined top altermagnetic layer ($d_{t}\!=\!2$ nm) and bottom inverse altermagnetic layer ($d_{b}\!=\!2$ nm), which gap both surface Dirac cones. Here, $G(z)\cos0\!=\!F(z)\!=\!1$ for $0\!\leqslant\!z\!\leqslant\!2$ nm, $G(z)\cos\pi\!=\!-F(z)\!=\!-1$ for $8\!\leqslant\!z\!\leqslant\!10$ nm, and $G(z)\!=\!F(z)\!=\!0$ elsewhere.
(d1) Top and bottom altermagnetic layers with parallel N\'{e}el vectors, $d_{t}\!=\!d_{b}\!=\!2$ nm and $\theta\!=\!0$, simultaneously gap both surfaces. Here, $G(z)\cos0\!=\!F(z)\!=\!1$ for $0\!\leqslant\!z\!\leqslant\!2$ nm and $8\!\leqslant\!z\!\leqslant\!10$ nm, and $G(z)\!=\!F(z)\!=\!0$ elsewhere.
Yellow shaded regions indicate the bandwidths of the corresponding surface states.
[(a2)-(d2)] Hall conductance as a function of the Fermi energy $E_{F}$ for the configurations in (a1)-(d1), exhibiting positive half-quantized, negative half-quantized, vanishing, and integer-quantized values in the energy gaps, respectively. In (c2), red (blue) dots represent the summed layer Hall conductance of the top (bottom) two layers.
Other parameters: $M\!=\!0.28$ eV, $A_{1}\!=\!0.22$ eV$\cdot$nm, $A_{2}\!=\!0.41$ eV$\cdot$nm, $B_{1}\!=\!0.1$ eV$\cdot$nm$^2$, $B_{2}\!=\!0.566$ eV$\cdot$nm$^2$, $\Delta\!=\!0.1$ eV, $J_{d}\!=\!B_{2}$, $\varphi\!=\!0$, $a_{z}\!=\!a_{||}\!=\!1$ nm, and sample thickness $L_{z}\!=\!10$ nm along the $z$ direction.
} \label{Fig:E_C_JB2_D01_Lz10_n2_inverse_together}
\end{figure*}

We now present numerical results for the $d$-wave altermagnets. The altermagnetic strength is chosen to be of the same order as the quadratic term in Bi$_2$Se$_3$, i.e., $J_{d} \!=\! B_{2}$.

\subsection{Tight-binding Hamiltonian}\label{3.1}

To perform the numerical calculations, we map the continuum Hamiltonian \eqref{eq:Hk_original} onto its tight-binding form in momentum space (see Sec. SII A of the Supplemental Material~\cite{SuppMat}):
\begin{eqnarray}\label{eq:Hk_TB}
\hat{\cal H}_{\rm TB}^{}({\bf k}) \!=\! \hat{\cal H}_{\rm TI}^{\rm TB}({\bf k}) \!+\! \hat{\cal H}_{\Delta}^{} \!+\! \hat{\cal H}_{J}^{\rm TB}({\bf k}_{||}),
\end{eqnarray} where 
\begin{eqnarray}
\hat{\cal H}_{\rm TI}^{\rm TB}({\bf k}) 
&\!=\!&{\cal M}_{\rm TB}^{}({\bf k})\sigma_{0}^{}\otimes\tau_{z}\!+\!\lambda_{z}\sin(k_{z}a_{z})\sigma_{0}^{}\otimes\tau_{y} \nonumber\\
&&\!+ \lambda_{||}\sin(k_{y}a_{||})\sigma_{x}\otimes\tau_{x} \nonumber\\
&&\!- \lambda_{||}\sin(k_{x}a_{||})\sigma_{y}\otimes\tau_{x},\label{eq:Htb_TB}\\
{\cal M}_{\rm TB}^{}({\bf k})
&\!=\!&(M \!-\! 2t_{z} \!-\! 4t_{||}^{}) \!+\! 2t_{z}\cos(k_{z}a_{z}) \nonumber\\
&&\!+ 2t_{||}^{}\left[\!\cos(k_{x}a_{||}^{}) \!+\! \cos(k_{y}a_{||}^{})\!\right]\!\!,\label{eq:M_k}\\
\hat{\cal H}_{J}^{\rm TB}({\bf k}_{||}) &\!=\!& 2G(z)J_{||}\cos\theta\!\!\left[\! \cos(k_{x}a_{||}^{}) \!-\! \cos(k_{y}a_{||}^{})\!\right]\!\!\sigma_{z}\otimes\tau_{0}^{}.\label{eq:Htb_Jd} \nonumber\\
\end{eqnarray} The parameters are defined as $t_{z}\!=\!B_{1}/a_{z}^2$, $t_{||}^{}\!=\!B_{2}/a_{||}^{2}$, $\lambda_{z}\!=\!A_{1}/a_{z}$, $\lambda_{||}\!=\!A_{2}/a_{||}^{}$, and $J_{||}\!=\!J_{d}/a_{||}^{2}$, with $a_{x}\!=\!a_{y}\!=\!a_{||}^{}$ denoting the lattice constants. 

Under open boundary conditions (OBCs) along the $z$ direction and periodic boundary conditions (PBCs) along $x$ and $y$ directions, the real-space tight-binding Hamiltonian in the basis 
$(\hat{C}_{k_x,k_y,1}^{}, \hat{C}_{k_x,k_y,2}^{},\hat{C}_{k_x,k_y,3}^{},\cdots,\hat{C}_{k_x,k_y,N_z}^{})^{T}$ is given by  (see Sec. SII B of the Supplemental Material~\cite{SuppMat})
\begin{eqnarray}\label{eq:Hz_TB_r}
\hat{\cal H}_{\rm TB}^{}({\bf k}_{||})\!=\!\begin{pmatrix}
\hat{h}({\bf k}_{||}) & \hat{T}_{z} & 0 & \cdots & 0 \\
\hat{T}_{z}^{\dagger} & \hat{h}({\bf k}_{||}) & \hat{T}_{z} & \cdots & 0 \\
0 & \hat{T}_{z}^{\dagger} & \hat{h}({\bf k}_{||}) & \ddots & \vdots \\
\vdots& \ddots & \ddots & \ddots & \hat{T}_{z} \\
0 & \cdots & 0 & \hat{T}_{z}^{\dagger} & \hat{h}({\bf k}_{||})
\end{pmatrix}_{(4N_z)\times (4N_z)}, \nonumber\\
\end{eqnarray} where
\begin{eqnarray}
\hat{h}({\bf k}_{||})
&\!=\!&\hat{M}_{0} \!+\! \hat{T}_{x}e^{ik_{x}a_{||}^{}} \!+\! \hat{T}_{x}^{\dagger}e^{-ik_{x}a_{||}^{}} \nonumber\\&&\!+ \hat{T}_{y}e^{ik_{y}a_{||}^{}} \!+\! \hat{T}_{y}^{\dagger}e^{-ik_{y}a_{||}^{}}, \\
\hat{M}_{0}&\!=\!&\left(M \!-\! 2t_{z} \!-\! 4t_{||}^{}\right)\sigma_{0}^{}\otimes\tau_{z} \nonumber\\
&&\!+ F(z)(\Delta_{x}\sigma_{y} \!-\! \Delta_{y}\sigma_{x})\otimes\tau_{0}^{},\label{eq:Dirac_mass_0}\\
\hat{T}_{z}&\!=\!&t_{z}\sigma_{0}^{}\otimes\tau_{z}\!-\! i\frac{\lambda_{z}}{2}\sigma_{0}^{}\otimes\tau_{y},\\
\hat{T}_{z}^{\dagger}&\!=\!&t_{z}\sigma_{0}^{}\otimes\tau_{z} \!+\! i\frac{\lambda_{z}}{2}\sigma_{0}^{}\otimes\tau_{y},
\end{eqnarray} $\hat{T}_{x}\!=\!t_{||}^{}\sigma_{0}^{}\otimes\tau_{z} \!+\! i\frac{\lambda_{||}}{2}\sigma_{y}\otimes\tau_{x} \!+\! G(z)J_{||}\cos\theta\sigma_{z}\otimes\tau_{0}^{}$, 
$\hat{T}_{x}^{\dagger}\!=\!t_{||}\sigma_{0}^{}\otimes\tau_{z} \!-\! i\frac{\lambda_{||}}{2}\sigma_{y}\otimes\tau_{x} \!+\! G(z)J_{||}\cos\theta\sigma_{z}\otimes\tau_{0}^{}$, 
$\hat{T}_{y}\!=\!t_{||}\sigma_{0}^{}\otimes\tau_{z}\!-\! i\frac{\lambda_{||}}{2}\sigma_{x}\otimes\tau_{x} \!-\! G(z)J_{||}\cos\theta\sigma_{z}\otimes\tau_{0}^{}$, and 
$\hat{T}_{y}^{\dagger}\!=\!t_{||}\sigma_{0}^{}\otimes\tau_{z} \!+\! i\frac{\lambda_{||}}{2}\sigma_{x}\otimes\tau_{x} \!-\! G(z)J_{||}\cos\theta\sigma_{z}\otimes\tau_{0}^{}$.

\subsection{Topological phases}\label{3.2}

Figure~\ref{Fig:E_C_JB2_D01_Lz10_n2_inverse_together} displays the band structures and Hall conductances for the tight-binding Hamiltonian \eqref{eq:Hz_TB_r} of Bi$_2$Se$_3$ coupled to $d$-wave altermagnetic layers under an external in-plane magnetic field. To elucidate the distinct roles of the altermagnet and the external field, we summarize the resulting topological phases below, referring directly to Fig.~\ref{Fig:E_C_JB2_D01_Lz10_n2_inverse_together}.

\begin{itemize}
\item Figures~\ref{Fig:E_C_JB2_D01_Lz10_n2_inverse_together}(a1) and \ref{Fig:E_C_JB2_D01_Lz10_n2_inverse_together}(a2): Altermagnet-induced half-quantized Hall effect. The top-surface Dirac cone is gapped by the altermagnet, while the bottom-surface Dirac cone remains gapless [Fig.~\ref{Fig:E_C_JB2_D01_Lz10_n2_inverse_together}(a1)]. This yields a positive half-quantized Hall conductance [Fig.~\ref{Fig:E_C_JB2_D01_Lz10_n2_inverse_together}(a2)], with the plateau width determined by the surface gap. Here, $\theta\!=\!0$, $F(z)\!=\!G(z)\!=\!1$ for $0\!\leqslant\!z\!\leqslant\!2$ nm and $F(z)\!=\!G(z)\!=\!0$ elsewhere. The sample thickness along the $z$ direction is $L_{z}\!=\!10$ nm.

\item Figures~\ref{Fig:E_C_JB2_D01_Lz10_n2_inverse_together}(b1) and \ref{Fig:E_C_JB2_D01_Lz10_n2_inverse_together}(b2): Altermagnet-induced half-quantized Hall effect (opposite sign). When the altermagnet acts only on the bottom layers [$\theta\!=\!\pi$ and $G(z)\!=\!1$], the bottom-surface Dirac cone is gapped while the top surface remains gapless [Fig.~\ref{Fig:E_C_JB2_D01_Lz10_n2_inverse_together}(b1)]. This configuration produces a negative half-quantized Hall conductance [Fig.~\ref{Fig:E_C_JB2_D01_Lz10_n2_inverse_together}(b2)], again with a plateau width determined by the gapped bandwidth.
The sign reversal of the Hall conductance arises from $G(z)\cos\theta\!=\!-1$ with $\theta\!=\!\pi$ and $G(z)\!=\!1$ in the bottom layers, contrasting with the positive value in Fig.~\ref{Fig:E_C_JB2_D01_Lz10_n2_inverse_together}(a2).

\item Figures~\ref{Fig:E_C_JB2_D01_Lz10_n2_inverse_together}(c1) and \ref{Fig:E_C_JB2_D01_Lz10_n2_inverse_together}(c2): Altermagnet-induced layer Hall effect. Both the top and bottom Dirac cones are gapped by altermagnets with antiparallel N\'{e}el vectors [Fig.~\ref{Fig:E_C_JB2_D01_Lz10_n2_inverse_together}(c1)], with $G(z)\cos0\!=\!F(z)\!=\!1$ for $0\!\leqslant\!z\!\leqslant\!2$ nm and $G(z)\cos\pi\!=\!-F(z)\!=\!-1$ for $8\!\leqslant\!z\!\leqslant\!10$ nm. The resulting Hall conductances on the two surfaces cancel, giving zero net Hall response, as seen in Fig.~\ref{Fig:E_C_JB2_D01_Lz10_n2_inverse_together}(c2). Red (blue) dots indicate the summed Hall conductance from the top (bottom) two layers, and the width of the quantized plateau of the summed layer Hall conductance is determined by the gapped bandwidth.

\item Figures~\ref{Fig:E_C_JB2_D01_Lz10_n2_inverse_together}(d1) and \ref{Fig:E_C_JB2_D01_Lz10_n2_inverse_together}(d2):
Altermagnet-induced anomalous Hall effect (Chern insulator). When both the top and bottom altermagnetic layers have the same sign of $G(z)$, both Dirac cones are gapped with identical sign [Fig.~\ref{Fig:E_C_JB2_D01_Lz10_n2_inverse_together}(d1)], resulting in a quantized Chern insulating state with total Hall conductance $e^{2}/h$ [Fig.~\ref{Fig:E_C_JB2_D01_Lz10_n2_inverse_together}(d2)], and the width of the quantized plateau is determined by the gapped bandwidth.
\end{itemize}

In Figs.~\ref{Fig:E_C_JB2_D01_Lz10_n2_inverse_together}(a1)-\ref{Fig:E_C_JB2_D01_Lz10_n2_inverse_together}(d1), the numerical spectra (black curves) obtained under OBCs along $z$ direction and PBCs along $x$, $y$ directions agree with the analytical surface-state dispersions [Eqs.~\eqref{eq:Ek_top} and \eqref{eq:Ek_bot}], shown in red (top surface) and blue (bottom surface). The bands are labeled as \lq\lq$t+$\rq\rq{} (top-surface conduction), \lq\lq$t-$\rq\rq{} (top-surface valence), \lq\lq$b+$\rq\rq{} (bottom-surface conduction), and \lq\lq$b-$\rq\rq{} (bottom-surface valence).


\section{Surface states}\label{4}

In this section, we present the effective Hamiltonians and corresponding eigenenergies for the surface states, and analytically derive the associated Hall conductances.

The effective Hamiltonian for the top surface state is given by (see Section SIII C of the Supplemental Material~\cite{SuppMat}),
\begin{eqnarray}
\hat{\cal H}_{\rm sur}^{\rm top}({\bf k}_{||})&\!=\!&-A_{2}\!\left(\!k_{y} \!+\! \frac{\Delta_{y}}{A_{2}}\!\right)\!\sigma_{x}\!+\!A_{2}\!\left(\!k_{x} \!+\! \frac{\Delta_{x}}{A_{2}}\!\right)\!\sigma_{y} \nonumber\\
&&\!+ J_{\rm top}(k_{x},k_{y})\sigma_{z}.
\label{eq:H_top}
\end{eqnarray} 
The corresponding eigenenergies for the top surface states are 
\begin{eqnarray}
E_{\text{sur}}^{{\rm top}(\pm)}({\bf k}_{||})
\!=\!\pm\sqrt{\!A_{2}^{2}\left[E_{0}^{\rm top}(k_{x},k_{y})\right]^{2}\!+\!\left[J_{\rm top}(k_{x},k_{y})\right]^{2}},\label{eq:Ek_top}\nonumber\\
\end{eqnarray} where $\left[E_{0}^{\rm top}(k_{x},k_{y})\right]^{2}\!=\!\left(\!k_{y} \!+\! \frac{\Delta_{y}}{A_{2}}\!\right)^{2}\!\!+\!\left(\!k_{x} \!+\! \frac{\Delta_{x}}{A_{2}}\!\right)^{2}$.

Similarly, the effective Hamiltonian for the bottom surface state is obtained as (see Section SIII C of the Supplemental Material~\cite{SuppMat})
\begin{eqnarray}
\hat{\cal H}_{\rm sur}^{\rm bot}({\bf k}_{||})&\!=\!&A_{2}\!\left(\!k_{y} \!-\! \frac{\Delta_{y}}{A_{2}}\!\right)\!\sigma_{x}\!-\!A_{2}\!\left(\!k_{x} \!-\! \frac{\Delta_{x}}{A_{2}}\!\right)\!\sigma_{y}\nonumber\\
&&\!+ J_{\rm bot}(k_{x},k_{y})\sigma_{z}.
\label{eq:H_bot}
\end{eqnarray} 
The corresponding eigenenergies are
\begin{eqnarray}
E_{\text{sur}}^{{\rm bot}(\pm)}({\bf k}_{||})
\!=\!\pm\sqrt{\!A_{2}^{2}\left[E_{0}^{\rm bot}(k_{x},k_{y})\right]^{2}\!+\!\left[J_{\rm bot}(k_{x},k_{y})\right]^{2}},\label{eq:Ek_bot}\nonumber\\
\end{eqnarray} where $\left[E_{0}^{\rm bot}(k_{x},k_{y})\right]^{2}\!=\!\left(\!k_{y} \!-\! \frac{\Delta_{y}}{A_{2}}\!\right)^{2}\!\!+\!\left(\!k_{x} \!-\! \frac{\Delta_{x}}{A_{2}}\!\right)^{2}$.

\subsection{Hall conductance versus magnetic-field orientation}\label{4.1}

\begin{figure}[htpb]
\centering
\includegraphics[width=\columnwidth]{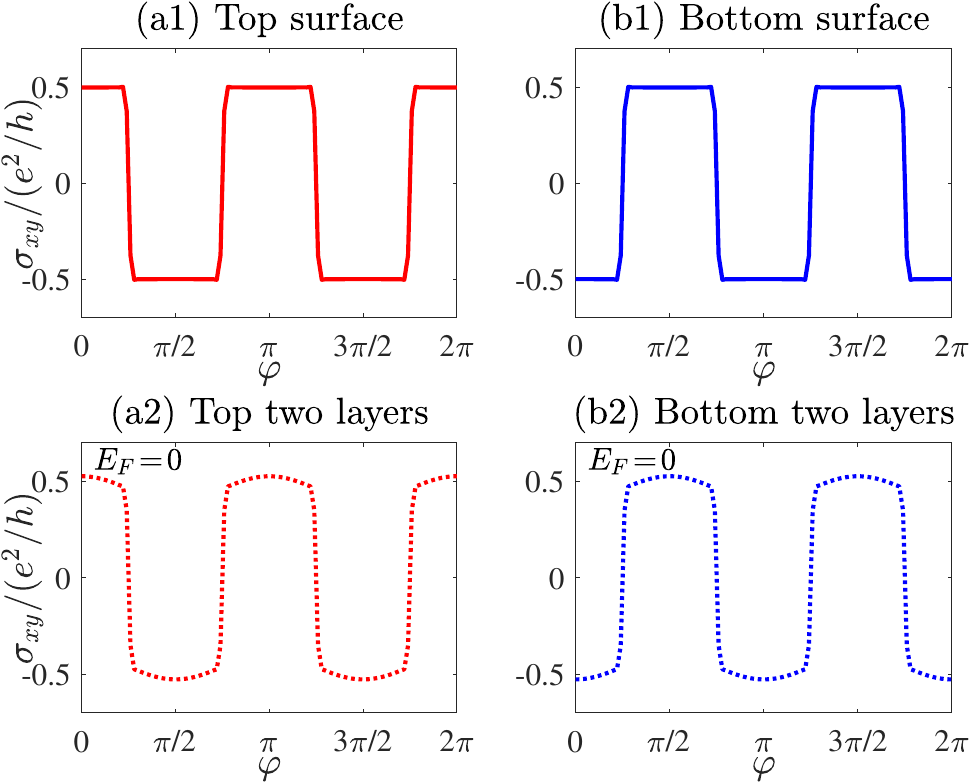}
\caption{Upper row: [(a1), (b1)] Hall conductances of the top and bottom surfaces as functions of the angle $\varphi$, calculated from the analytical expressions in Eqs.~\eqref{eq:Hallconductance_top_1} and \eqref{eq:Hallconductance_bot_1}, respectively. (a1) Top-surface Hall conductance with $\theta\!=\!0$ and $G(z)\cos0\!=\!F(z)\!=\!1$. (b1) Bottom-surface Hall conductance with $\theta\!=\!\pi$ and $G(z)\cos\pi\!=\!-F(z)\!=\!-1$. Here, $\varphi$ specifies the orientation of the external in-plane magnetic field. The integration range is $k_{x},k_{y}\in[-\pi,\pi]$ nm$^{-1}$, and the Fermi energy $E_{F}$ is set within the surface-state gap.
Lower row: [(a2), (b2)] Summed layer Hall conductance of the top (bottom) two layers as a function of $\varphi$, for a system with both a top altermagnetic layer ($d_{t}\!=\!2$ nm) and a bottom inverse altermagnetic layer ($d_{b}\!=\!2$ nm). The spatial profiles are given by $G(z)\cos0\!=\!F(z)\!=\!1$ for $0\!\leqslant\!z\!\leqslant\!2$ nm, $G(z)\cos\pi\!=\!-F(z)\!=\!-1$ for $8\!\leqslant\!z\!\leqslant\!10$ nm, and $G(z)\!=\!F(z)\!=\!0$ elsewhere. (a2) Summed Hall conductance of the top two layers. (b2) Summed Hall conductance of the bottom two layers. Here, $E_{F}\!=\!0$, and all other parameters are the same as in Fig.~\ref{Fig:E_C_JB2_D01_Lz10_n2_inverse_together}.
} \label{Fig:C_layer_JB2_D01_Lz10_inverse_varphi_together}
\end{figure}

\begin{figure*}[htpb]
\centering
\includegraphics[width=\textwidth]{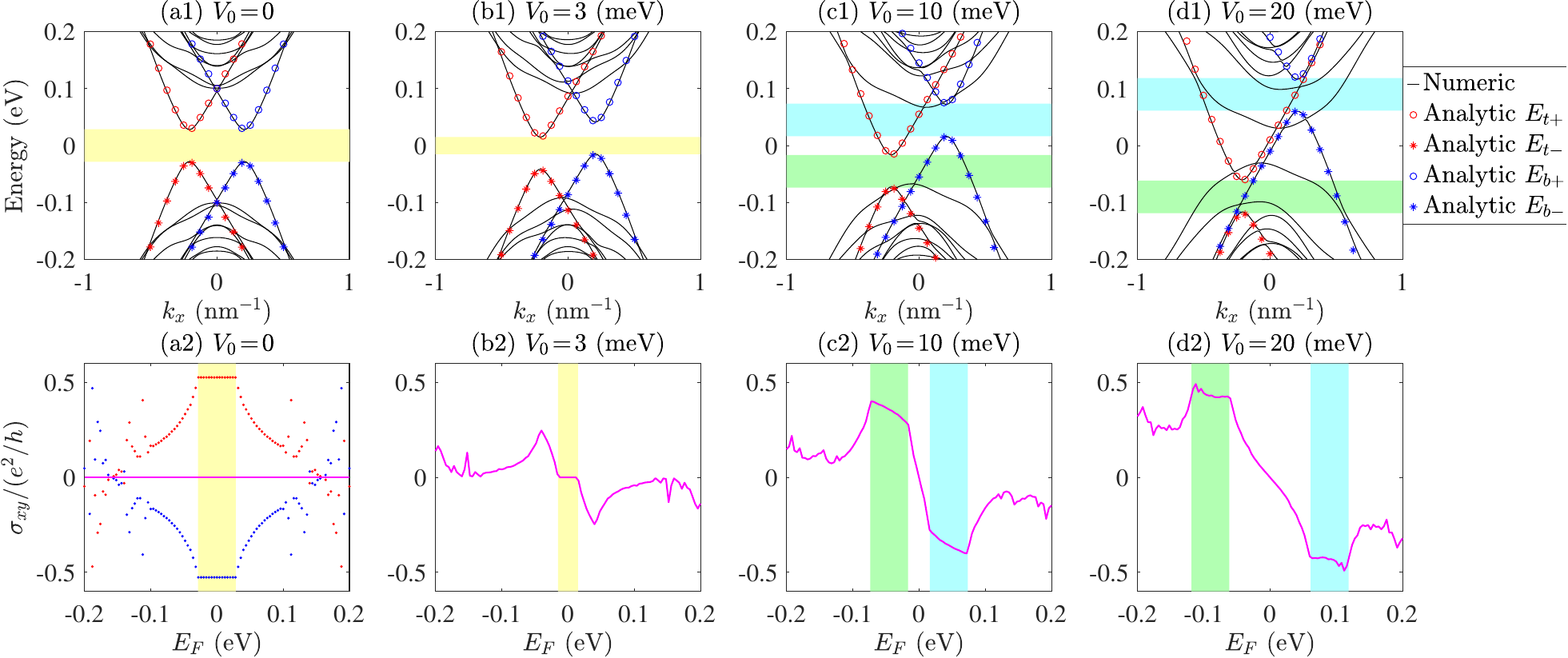}
\caption{Band structures and Hall conductances for the altermagnet-induced layer Hall effect under different strengths $V_{0}$, which could be induced by applying the perpendicular electric field.
Upper row: [(a1)-(d1)] show the energy spectra for (a1) $V_{0}\!=\!0$, (b1) $V_{0}\!=\!3$ meV, (c1) $V_{0}\!=\!10$ meV, and (d1) $V_{0}\!=\!20$ meV. Black curves show numerical results under OBCs along $z$ direction and PBCs along $x$ and $y$ directions. Red (blue) circles and asterisks represent analytical surface-state spectra for the top (bottom) surface obtained from Eqs.~\eqref{eq:Ek_top_Ez} and \eqref{eq:Ek_bot_Ez}. In (a1) and (b1), the yellow shaded region marks the bandwidth of the surface states, while in (c1) and (d1), the cyan (green) shaded regions indicate the bandwidths of the top (bottom) surface states. 
Lower row: [(a2)-(d2)] display the corresponding Hall conductance as a function of the Fermi energy $E_{F}$. Magenta lines show the total Hall conductance, $\sigma_{xy}\!=\!\sum_{j_z}\sigma_{xy}(j_z)$. In (a2), red (blue) dots denote the summed layer Hall conductance of the top (bottom) two layers. 
The system consists of a top altermagnetic layer ($d_{t}\!=\!2$ nm) and a bottom inverse altermagnetic layer ($d_{b}\!=\!2$ nm), with $G(z)\cos0\!=\!F(z)\!=\!1$ for $0\!\leqslant\!z\!\leqslant\!2$ nm, $G(z)\cos\pi\!=\!-F(z)\!=\!-1$ for $8\!\leqslant\!z\!\leqslant\!10$ nm, and $G(z)\!=\!F(z)\!=\!0$ elsewhere. These parameters are identical to those used in Figs.~\ref{Fig:E_C_JB2_D01_Lz10_n2_inverse_together}(c1) and \ref{Fig:E_C_JB2_D01_Lz10_n2_inverse_together}(c2). All other parameters are the same as those used in Fig.~\ref{Fig:E_C_JB2_D01_Lz10_n2_inverse_together}.
} \label{Fig:E_C_JB2_D01_Lz10_n2_inverse_together_Vz03}
\end{figure*}

Based on the effective Hamiltonians \eqref{eq:H_top} and \eqref{eq:H_bot}, we consider a $d$-wave altermagnetic term $\hat{\cal H}_{J}^{}({\bf k}_{||})\!=\!G(z)J_{d}(k_{y}^{2} \!-\! k_{x}^{2})\cos\theta\sigma_{z}\otimes\tau_{0}^{}$~\cite{smejkal2022giant,smejkal2022beyond,smejkal2022emerging}. The Hall conductances for the surface states can then be obtained analytically as (see Section SIV of the Supplemental Material~\cite{SuppMat})
\begin{eqnarray}
\sigma_{xy}^{\rm top}
&\!=\!&\frac{e^{2}}{2h}\!\!\int\!\!\frac{J_{+}G(z)\cos\theta}{\left\{d_{+}^{2} \!+\! J_{\rm sur}^{2}[G(z)\cos\theta]^{2}\right\}^{3/2}} \frac{dk_{x}dk_{y}}{2\pi}, \label{eq:Hallconductance_top_1} \\
\sigma_{xy}^{\rm bot}
&\!=\!&\frac{e^{2}}{2h}\!\!\int\!\!\frac{J_{-}G(z)\cos\theta}{\left\{d_{-}^{2} \!+\! J_{\rm sur}^{2}[G(z)\cos\theta]^{2}\right\}^{3/2}} \frac{dk_{x}dk_{y}}{2\pi}, \label{eq:Hallconductance_bot_1}
\end{eqnarray}
where $J_{\pm}\!=\!A_{2}^{2}J_{d}\left[k_{y}\left(k_{y} \!\pm\! 2\frac{\Delta_{y}}{A_{2}} \right) \!-\! k_{x}\left(k_{x} \!\pm\! 2\frac{\Delta_{x}}{A_{2}} \right)\right]$, $d_{\pm}\!=\!\left\{A_{2}^{2}\left[\left( k_{y} \!\pm\! \frac{\Delta_{y}}{A_{2}} \right)^{2} \!+\! \left( k_{x} \!\pm\! \frac{\Delta_{x}}{A_{2}} \right)^{2}\right]\right\}^{1/2}$, $J_{\rm sur}\!=\!J_{d}\left( k_{y}^{2} \!-\! k_{x}^{2} \right)$, $k_{x},k_{y}\in[-\pi,\pi]$ nm$^{-1}$, and the Fermi energy $E_{F}$ is assumed to lie within the energy gap of the surface states.

We further explore the dependence of the Hall conductance on the orientation $\varphi$ of the external in-plane magnetic field in the case of the altermagnet-induced layer Hall effect. Figures~\ref{Fig:C_layer_JB2_D01_Lz10_inverse_varphi_together}(a1) and \ref{Fig:C_layer_JB2_D01_Lz10_inverse_varphi_together}(b1) show the analytically calculated Hall conductances of the top and bottom surfaces, based on Eqs.~\eqref{eq:Hallconductance_top_1} and \eqref{eq:Hallconductance_bot_1}, with the Fermi energy $E_{F}$ lying within the surface gap. The sign of the half-quantized Hall conductance is tunable by varying $\varphi$, and the top and bottom surfaces exhibit opposite signs as expected for the layer Hall effect.

For comparison, the numerically evaluated summed Hall conductances of the top and bottom two layers at $E_{F}\!=\!0$ are plotted in Figs.~\ref{Fig:C_layer_JB2_D01_Lz10_inverse_varphi_together}(a2) and \ref{Fig:C_layer_JB2_D01_Lz10_inverse_varphi_together}(b2), respectively. The numerical results are in good agreement with the analytical predictions, confirming the robustness of the altermagnet-induced layer Hall effect against variations in field orientation.
Interestingly, the Hall conductance of the top (bottom) surface states exhibits periodic sign reversals at $\varphi \approx \frac{(2n\!+\!1)\pi}{4}$ ($n\!=\!0,1,2,3$).

\section{Perpendicular electric field for layer Hall effect}\label{5}

In this section, we demonstrate that the layer Hall effect becomes experimentally observable upon applying a perpendicular (out-of-plane) electric field $E_{z}$~\cite{gao2021layer, chen2024layer}.

We assume that the electric field $E_{z}$ induces a potential $\hat{V}(z)$ that is an odd function of the out-of-plane coordinate $z$, expressed as~\cite{gao2021layer, chen2024layer}
\begin{eqnarray}\label{eq:Vz}
\hat{V}(z)\!=\!V_{0}\!\left[j_{z} \!-\! \frac{1}{2}(N_{z}\!+\!1)\right]\!\sigma_{0}^{}\otimes\tau_{0}^{},
\end{eqnarray}
where $V_{0}$ characterizes the layer-dependent potential induced by $E_{z}$, and $j_{z} \!=\! 1, 2, 3, \cdots, N_{z}$  indexes the individual layers with $j_{z}\!=\!z/a_{z}$ and $N_{z}\!=\!L_{z}/a_{z}$. Here, $L_{z}$ denotes the sample thickness along the $z$ direction.

Incorporating this potential $\hat{V}(z)$ into the real-space tight-binding Hamiltonian~\eqref{eq:Hz_TB_r}, the Dirac mass term in Eq.~\eqref{eq:Dirac_mass_0} is modified as 
\begin{eqnarray}
\hat{M}(z)&\!=\!&\hat{M}_{0} \!+\! \hat{V}(z) \nonumber\\
&\!=\!&\left(M \!-\! 2t_{z} \!-\! 4t_{||}\right)\sigma_{0}^{}\otimes\tau_{z} \nonumber\\
&&\!+ F(z)(\Delta_{x}\sigma_{y} \!-\! \Delta_{y}\sigma_{x})\otimes\tau_{0}^{} \!+\! \hat{V}(z).\label{eq:Dirac_mass_Ez}
\end{eqnarray} 

To analytically investigate the effect of $E_{z}$ on the surface states, we derive the effective surface Hamiltonians for both the top and bottom surfaces.

On the basis of the potential $\hat{V}(z)$, i.e., Eq.~\eqref{eq:Vz}, the effective Hamiltonian for the top surface state reads
\begin{eqnarray}
\hat{\cal H}_{\rm sur}^{\rm top}({\bf k}_{||})&\!=\!& - A_{2}\!\left(\!k_{y} \!+\! \frac{\Delta_{y}}{A_{2}}\!\right)\!\sigma_{x}\!+\!A_{2}\!\left(\!k_{x} \!+\! \frac{\Delta_{x}}{A_{2}}\!\right)\!\sigma_{y} \nonumber\\
&&\!+ J_{\rm top}(k_{x},k_{y})\sigma_{z} \!+\! \frac{1}{2}(1 \!-\! N_{z})V_{0}\sigma_{0}^{}\otimes\tau_{0}^{}.\nonumber\\
\label{eq:H_top_Ez}
\end{eqnarray} 
The corresponding eigenenergies for the top surface states are given by
\begin{eqnarray}
E_{\text{sur}}^{{\rm top}(\pm)}({\bf k}_{||})
&\!=\!&\frac{1}{2}(1 \!-\! N_{z})V_{0} \nonumber\\
&&\pm\sqrt{\!A_{2}^{2}\left[E_{0}^{\rm top}(k_{x},k_{y})\right]^{2}\!+\!\left[J_{{\rm top}}(k_{x},k_{y})\right]^{2}},\label{eq:Ek_top_Ez}\nonumber\\
\end{eqnarray} where $\left[E_{0}^{\rm top}(k_{x},k_{y})\right]^{2}\!=\!\left(\!k_{y} \!+\! \frac{\Delta_{y}}{A_{2}}\!\right)^{2}\!\!+\!\left(\!k_{x} \!+\! \frac{\Delta_{x}}{A_{2}}\!\right)^{2}$.

Furthermore, the effective Hamiltonian for the bottom surface state is evaluated as
\begin{eqnarray}
\hat{\cal H}_{\rm sur}^{\rm bot}({\bf k}_{||})&\!=\!&A_{2}\!\left(\!k_{y} \!-\! \frac{\Delta_{y}}{A_{2}}\!\right)\!\sigma_{x}\!-\!A_{2}\!\left(\!k_{x} \!-\! \frac{\Delta_{x}}{A_{2}}\!\right)\!\sigma_{y}\nonumber\\
&&\!+ J_{\rm bot}(k_{x},k_{y})\sigma_{z} \!+\! \frac{1}{2}(N_{z} \!-\! 1)V_{0}\sigma_{0}^{}\otimes\tau_{0}^{}.\nonumber\\
\label{eq:H_bot_Ez}
\end{eqnarray} 
The corresponding eigenenergies for the bottom surface states are given by
\begin{eqnarray}
E_{\text{sur}}^{{\rm bot}(\pm)}({\bf k}_{||})
&\!=\!&\frac{1}{2}(N_{z} \!-\! 1)V_{0} \nonumber\\
&&\pm\sqrt{\!A_{2}^{2}\left[E_{0}^{\rm bot}(k_{x},k_{y})\right]^{2}\!+\!\left[J_{\rm bot}(k_{x},k_{y})\right]^{2}},\label{eq:Ek_bot_Ez}\nonumber\\
\end{eqnarray} where $\left[E_{0}^{\rm bot}(k_{x},k_{y})\right]^{2}\!=\!\left(\!k_{y} \!-\! \frac{\Delta_{y}}{A_{2}}\!\right)^{2}\!\!+\!\left(\!k_{x} \!-\! \frac{\Delta_{x}}{A_{2}}\!\right)^{2}$.

As shown in Fig.~\ref{Fig:E_C_JB2_D01_Lz10_n2_inverse_together_Vz03}, we numerically calculate the band structures and Hall conductances for the altermagnet-induced layer Hall effect under different strengths $V_{0}$, which could be induced by applying the perpendicular electric field.

By comparing Figs.~\ref{Fig:E_C_JB2_D01_Lz10_n2_inverse_together_Vz03}(a1) and \ref{Fig:E_C_JB2_D01_Lz10_n2_inverse_together_Vz03}(b1), one observes that the bandwidth (yellow-shaded region) becomes significantly narrower upon applying a small electric field with $V_{0}\!=\!3$ meV. In the absence of $E_{z}$, the total Hall conductance $\sigma_{xy}\!=\!\sum_{j_z}\sigma_{xy}(j_z)$ (magenta curve) vanishes, as the opposite half-quantized layer Hall conductances from the top and bottom surfaces cancel each other [Fig.~\ref{Fig:E_C_JB2_D01_Lz10_n2_inverse_together_Vz03}(a2)]. However, when $E_{z}$ is applied, this cancellation only persists within the narrower yellow-shaded region [Fig.~\ref{Fig:E_C_JB2_D01_Lz10_n2_inverse_together_Vz03}(b2)], whose width is determined by the gapped bandwidth. Outside this region, the total Hall conductance becomes finite.

As $V_{0}$ increases further, the bulk gap gradually closes, lifting the exact cancellation between the two surfaces. The total Hall conductance then becomes finite and varies continuously with the Fermi energy $E_{F}$. Notably, two approximate plateaus appear, one positive (green) and one negative (cyan), as shown in Figs.~\ref{Fig:E_C_JB2_D01_Lz10_n2_inverse_together_Vz03}(c1) and \ref{Fig:E_C_JB2_D01_Lz10_n2_inverse_together_Vz03}(d1). The positive plateau corresponds to the gap of the top surface states, while the negative one arises from the bottom surface states. The widths of these plateaus are determined by the respective surface-state bandwidths.
The emergence of nonzero plateaus and peaks in the total Hall conductance thus provides clear and experimentally accessible signatures of the layer Hall effect.

\begin{figure}[htpb]
\centering
\includegraphics[width=\columnwidth]{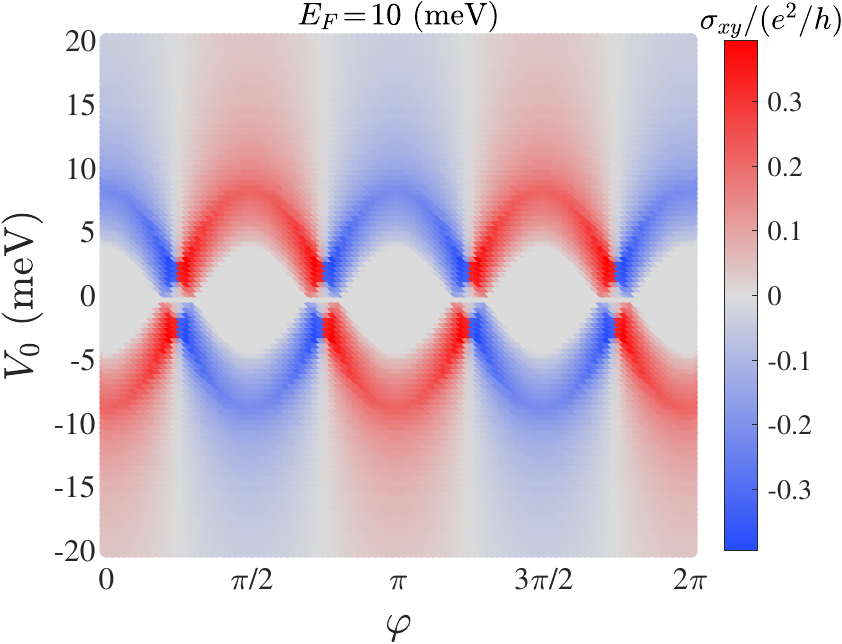}
\caption{Total Hall conductance as a function of $\varphi$ and $V_{0}$ at a fixed Fermi energy $E_{F}\!=\!10$ meV.
The system consists of a top altermagnetic layer ($d_{t}\!=\!2$ nm) and a bottom inverse altermagnetic layer ($d_{b}\!=\!2$ nm). The spatial profiles are specified as $G(z)\cos0\!=\!F(z)\!=\!1$ for $0\!\leqslant\!z\!\leqslant\!2$ nm, $G(z)\cos\pi\!=\!-F(z)\!=\!-1$ for $8\!\leqslant\!z\!\leqslant\!10$ nm, and $G(z)\!=\!F(z)\!=\!0$ elsewhere. These parameters are identical to those used in Figs.~\ref{Fig:E_C_JB2_D01_Lz10_n2_inverse_together}(c1) and \ref{Fig:E_C_JB2_D01_Lz10_n2_inverse_together}(c2). All other parameters are the same as those used in Fig.~\ref{Fig:E_C_JB2_D01_Lz10_n2_inverse_together}.
} \label{Fig:Phase_JB2_D01_Lz10_n2_top_bottom_inverse_Ef001_Vz}
\end{figure}

To explore the tunability of the total Hall conductance with respect to the orientation $\varphi$ of the external in-plane magnetic field and the perpendicular (out-of-plane) electric field, we plot the total Hall conductance as a function of $\varphi$ and $V_{0}$ at a fixed Fermi energy $E_{F}\!=\!10$ meV, as shown in Fig.~\ref{Fig:Phase_JB2_D01_Lz10_n2_top_bottom_inverse_Ef001_Vz}. When $V_{0}\!=\!0$, the total Hall conductance vanishes identically for all $\varphi$ due to the exact cancellation between the contributions from the top and bottom surface states. Upon applying a finite perpendicular electric field, this cancellation is lifted, giving rise to a finite Hall response. Particularly, the total Hall conductance undergoes periodic sign reversals at $\varphi \!\approx\! \frac{(2n\!+\!1)\pi}{4}$ ($n\!=\!0,1,2,3$), closely resembling the behavior of the individual surface Hall conductances shown in Fig.~\ref{Fig:C_layer_JB2_D01_Lz10_inverse_varphi_together}. Furthermore, when the electric field is applied symmetrically, i.e., $V_{0}\in[-20,20]$ meV, the total Hall conductance exhibits a $\pi$-periodic dependence on $\varphi$.

\section{Summary}\label{6}

In summary, we have proposed a scheme to realize the layer Hall effect in the ferromagnetic topological insulator Bi$_2$Se$_3$ through proximity to $d$-wave altermagnets. We demonstrated that the combination of an altermagnet and an in-plane magnetic field gaps the surface Dirac cone, leading to an altermagnet-induced half-quantized Hall effect. When altermagnets with antiparallel N\'{e}el vectors are applied to the top and bottom surfaces, producing a layer Hall effect with vanishing net Hall conductance. By contrast, when altermagnets with parallel N\'{e}el vectors are applied, yielding a quantized Chern insulating state, i.e., the altermagnet-induced anomalous Hall effect.

We further showed that the Hall conductance depends sensitively on the orientation of the in-plane magnetic field, providing an additional degree of control over the topological response. Moreover, we demonstrated that the layer Hall effect becomes experimentally observable upon applying a perpendicular electric field, which enables direct detection in realistic setups.

Unlike conventional systems, where the Hall response is typically controlled by uniform ferromagnetic order, our scheme leverages altermagnetic order combined with in-plane magnetic fields to induce surface-dependent Dirac gaps. This allows for the realization of a layer Hall effect with vanishing net Hall conductance, a feature that is not accessible in standard magnetic topological insulators. Additionally, the angular dependence of the layer Hall effect provides a direct probe of the $d$-wave symmetry of the altermagnet, offering an experimentally accessible way to detect and manipulate altermagnetic order. These features highlight the unique advantages of our approach in engineering topological phases beyond conventional axion insulator setups.

Taken together, these results establish a versatile framework for engineering altermagnet-induced topological phases in ferromagnetic topological insulators. Our findings not only deepen the understanding of altermagnetism in topological systems but also open avenues for the design and realization of altermagnet-based topological quantum materials.


\begin{acknowledgments}
R.C. acknowledges support from the National Natural Science Foundation of China (Grants No. 12304195 and No. U25D8012), the Chutian Scholars Program in Hubei Province, the Key Project of Hubei Provincial Department of Education (under Grant No. D20241004), the Hubei Provincial Natural Science Foundation (Grant No. 2025AFA081), the Wuhan City Key R\&D Program (under Grant No. 2025050602030069), and the Original Seed Program of Hubei University.
This work is supported by the Guangdong Provincial Quantum Science Strategic Initiative (Grant No. GDZX2401001). 
F.Q. acknowledges support from the Jiangsu Specially-Appointed Professor Program in Jiangsu Province and the Doctoral Research Start-Up Fund of Jiangsu University of Science and Technology. 
\end{acknowledgments}

\section*{Data Availability}
The data are available from the authors upon reasonable request.

%
%
%
\twocolumngrid
\bibliography{references_Altermagnets_Layer}

\end{document}